\documentclass[showpacs,amsmath,amssymb,twocolumn]{revtex4}
\pdfoutput=1 

\usepackage{graphicx}

\begin{document}
\title{Relativistic Quantum Cryptography}

\author{I. V. Radchenko}
\author{K. S. Kravtsov}
\email{kravtsov@kapella.gpi.ru}
\affiliation{A.M. Prokhorov General Physics Institute RAS, Moscow, Russia}
\author{S. P. Kulik}
\affiliation{Faculty of Physics, Moscow State University, Moscow, Russia}
\author{S. N. Molotkov}
\affiliation{Academy of Cryptography of Russian Federation, Moscow, Russia}
\affiliation{Institute of Solid State Physics, Chernogolovka, Moscow Rgn., Russia}
\affiliation{Faculty of Computational Mathematics and Cybernetics, Moscow State University, Moscow, Russia}

\begin{abstract}
Quantum key distribution (QKD) is a concept of secret key exchange supported by fundamentals of quantum physics. Its
perfect realization offers unconditional key security, however, known practical schemes are potentially vulnerable if
the quantum channel loss exceeds a certain realization-specific bound. This discrepancy is caused by the fact that any
practical photon source has a non-zero probability of emitting two or more photons at a time, while theory needs exactly
one. We report an essentially different QKD scheme based on both quantum physics and theory of relativity. It works
flawlessly with practical photon sources at arbitrary large channel loss. Our scheme is naturally tailored for
free-space optical channels, and may be used in ground-to-satellite communications, where losses are prohibitively large and
unpredictable for conventional QKD. 
\end{abstract}


\maketitle

Quantum cryptography~\cite{BB84,E91,GRT02,SBC09,HH11,LR13} gained its popularity from a promise of its absolute security
against eavesdropping. In this sense, ``absolute'' means that it is guaranteed by fundamental laws of physics,
rather than by our technological abilities. Nowadays, QKD is arguably the only practical technology explicitly operating
with properties of the quantum world. At the same time, { \em practical } QKD is a serious challenge for scientists,
because all implementations are somewhat different from underlying theoretical models. Two major problems are the lack
of true single-photon sources~\cite{LM00,B00} and the presence of loss in quantum channel; neither of them can be perfectly eliminated.
In result, known protocols guarantee key security only if losses do not exceed a certain realization-specific level. In
this paper we address this issue and demonstrate a {\em relativistic } protocol taking into account relativistic
properties of our world, which guarantee key security regardless of the particular channel loss.

Conventional, non-relativistic, quantum cryptography is based on fundamental principles of quantum
mechanics~\cite{D82,WZ82},
however, it is not
directly tied to elementary particles or other physical objects that carry transmitted quantum states. In the proposed
{\em relativistic quantum cryptography} it must be a massless particle traveling at the speed of light, i.e. a photon, that carries
information. This makes a difference
if the space-time structure of the communication in Minkowsky space is taken into account, calling to nonexistence of faster-than-light information
transmission. This explicit connection with the space-time is completely ignored in conventional QKD protocols.

Our protocol is based on time-spread coherent quantum states, which, due to their distributed nature
inevitably cause delays, if successful intercept and resend attack is performed. Thus, detection of adversary actions can
be performed by controlling both detection errors {\it and }signal delays. This eventually makes the protocol completely
immune to arbitrarily large loss in the quantum channel and creates a potential for its use in
ground-to-satellite free space quantum links enabling 
global QKD service~\cite{LR13}.

As signal delays play a critical role in our relativistic approach, the protocol is only viable for
line-of-sight free space communication links, where no alternative shortcut paths are possible and the signal propagates
at the speed of light. 
Importantly, the protocol is tolerant to the presence of air in the light path, which slightly delays the signal vs. the vacuum speed
of light; it only requires enough time-spreading of the transmitted quantum states. In typical conditions it requires about
1~ns of spreading per kilometer of the transmission distance in air, topping to less than 20~ns for ground-to-satellite links.

Although keeping track of precise timing requires, in general, external clock synchronization, the proposed protocol
takes care of clock synchronization between the communicating parties Alice
and Bob by itself; no other external synchronization scheme is needed.
At the same time the protocol requires an a-priori knowledge of
a distance between the parties, which is essential, e.g.  if the adversary Eve chooses to
delay any light transmissions between Alice and Bob.

The understanding that special relativity may offer new features to quantum cryptography, was around as early
as in 1990s, when the first {\it relativistic} quantum protocol on orthogonal states was proposed~\cite{GV95,P96,GV96,KI97,ABD10,XTW12}.
Another use of relativistic causality in QKD was demonstrated in~\cite{JAK06},
where a two-fold increase of key generation rate for BB84 protocol~\cite{BB84} was shown.

Even more substantial changes relativity provided for bit commitment protocols~\cite{K99,MN00,H11,K12,LKB13}, which are
prohibited in a conventional quantum world by the Mayers-Lo-Chau no-go theorems~\cite{M96,M97,LC97}.
The search for {\it qualitatively new} features in QKD was on since an early
work~\cite{MN01} in 2001. Followed by a series of publications, these ideas distilled into
a practical relativistic QKD protocol~\cite{M11,M11_VAR,M12}, experimentally demonstrated in the
present work for the first time.

The main features of the realized relativistic protocol include: (i) spreading of quantum states
does not have to be as large as the channel length; it only has to compensate for delays in the channel with respect to
the ideal one with the vacuum speed of light; (ii) as relativistic principles allow for clock
synchronization, no other external synchronization is required; (iii) the protocol provides unconditional key security even
with conventional faint laser pulses at arbitrarily large channel loss; practical limitations on the channel loss
are only determined by the dark count rate in the single photon detector used.

The realized system has a double-pass
configuration, where optical pulses initially generated by Bob are first transmitted to Alice as classical signals.
Alice in turn attenuates the received signals to quantum level, encodes them with randomly chosen bits and sends back to
Bob. Bob detects them at exactly calculated moments in time, choosing his measurement basis at random. Such
configuration allows reusing the same fiber delay interferometer, installed at Bob's,
and does not require its phase stabilization.

\begin{figure}
\centering
\includegraphics[width=\columnwidth]{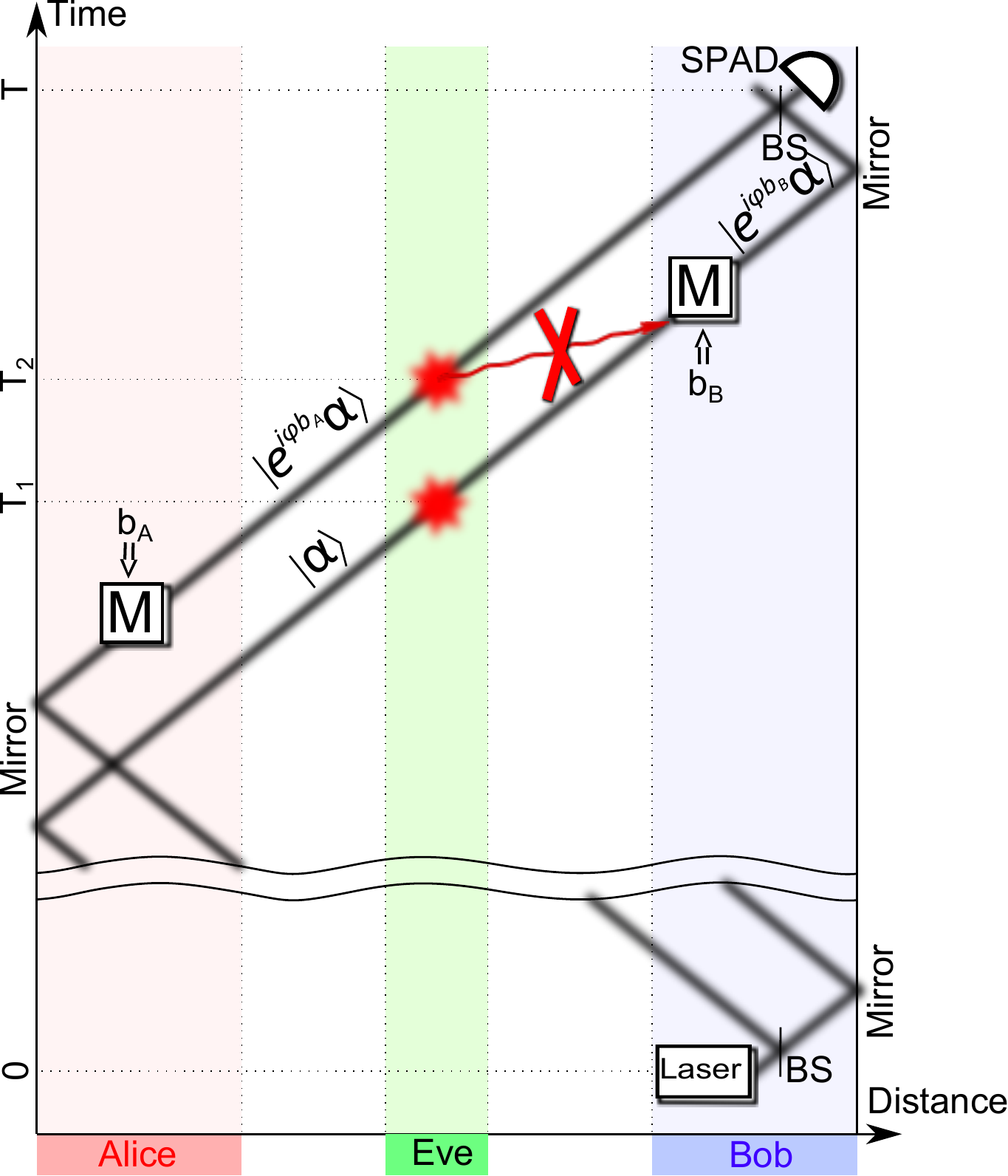}
\caption{{\bf Space-time diagram of the line-of-sight free-space quantum channel operation.} M --- phase modulator,
$|\alpha\rangle$ --- coherent quantum state with the average number of photons $|\alpha|^2 = \mu$, BS --- beam splitter,
SPAD --- single-photon avalanche detector. Each transmission is
initiated by Bob, who sends a classical laser pulse, split into two in the delay interferometer. Alice attenuates both
copies of the pulse to a quantum level, reflects them off the mirror, and modulates the phase of the second one with her
random bit of information $b_A$. Thus, the quantum transmission from Alice to Bob consists of a pair of coherent states,
which are then tested by Bob in the delay interferometer using phase modulation of the first state with his random bit
$b_B$. Inability of faster-than-light communication prevents Eve from successful eavesdropping because she cannot undo her
decision about the first pulse made at time $T_1$, when she gets the ``data'' pulse at the moment $T_2.$ Any attempts of
eavesdropping result in detection errors at the Bob's side or in additional signal delays, which are equivalent to
losses because Bob only detects signals in his short detection window at time $T$.}
\label{relativity} 
\end{figure}

The protocol serves for the two goals: first, to provide relativistic QKD by itself and, second, to
synchronize the clocks between Alice and Bob. The first goal is achieved by the following procedure shown in
figure~\ref{relativity} as a space-time diagram. Alice receives from
Bob pairs of short optical pulses spread in time by a fixed amount $\Delta t$. As each pair is generated in the delay
interferometer from a single laser pulse, the pair would interfere destructively in the detector port of the interferometer
if returned to Bob without any phase shifts. Alice flips a coin to decide whether to change the phase of the second
pulse by $\varphi$ or not. After performing (or not performing) the modulation she attenuates the pulses to achieve a
pre-defined average number of photons $\mu$ in each of the pulses and sends them back to Bob. Bob also randomly chooses
whether to apply an additional phase shift $\varphi$ to the first pulse or not and detects the result of their interference exactly
at the time $T$ after initial pulse generation. This value $T$ should be equal to the round-trip time with the accuracy
better than $\Delta t$. All time delays shorter than $\Delta t$ are considered insignificant.

As one may notice, so far the protocol directly repeats the B92 protocol~\cite{B92}, so where does relativity come into play? An
explanation is given by the following argument. There are two weak coherent states
in the channel: $|\alpha\rangle$ going first and $|e^{i\varphi\, b_A}\alpha\rangle$, where $b_A$ is the bit sent by
Alice and $\alpha = \sqrt{\mu} e^{i\varphi_0}$. The first state carries no information and is considered to be known by Eve. However, Bob will use it to test the
data state traveling later, thus, it is essential for the communication. If the first state gets blocked in the
transmission line, Bob's measurement result will not depend on his modulator setting, which will result in the 50\% error
rate. If, similarly, the second, data, state is missing, while the first is present, Bob apparently gets 50\% errors
again. 

Eve may choose whether to block or do not block the first state in the line, but when she
receives the second one, she faces uncertainty of the quantum world: as the data carrier states are not
orthogonal to each other $\left| \langle e^{i\varphi }\alpha | \alpha \rangle\right| = \exp[-2|\alpha|^2 \sin^2
(\varphi/2)]  > 0,$ it
is not possible to guarantee reconstruction of the bit sent. If Eve succeeds in obtaining the bit, e.g. with an unambiguous
measurement, she is okay only in the case if she has let the first state propagate in the line, because otherwise Bob
would see an error. If she fails to get the bit, she is okay only if the first state has been blocked. Due to
relativistic causality she cannot correct for her decision about the first pulse made earlier, as there is no faster-than-light
communication. Thus, introduction of errors into the line by Eve now does not depend on the channel loss.

\begin{figure}
\centering
\includegraphics[width=0.8\columnwidth]{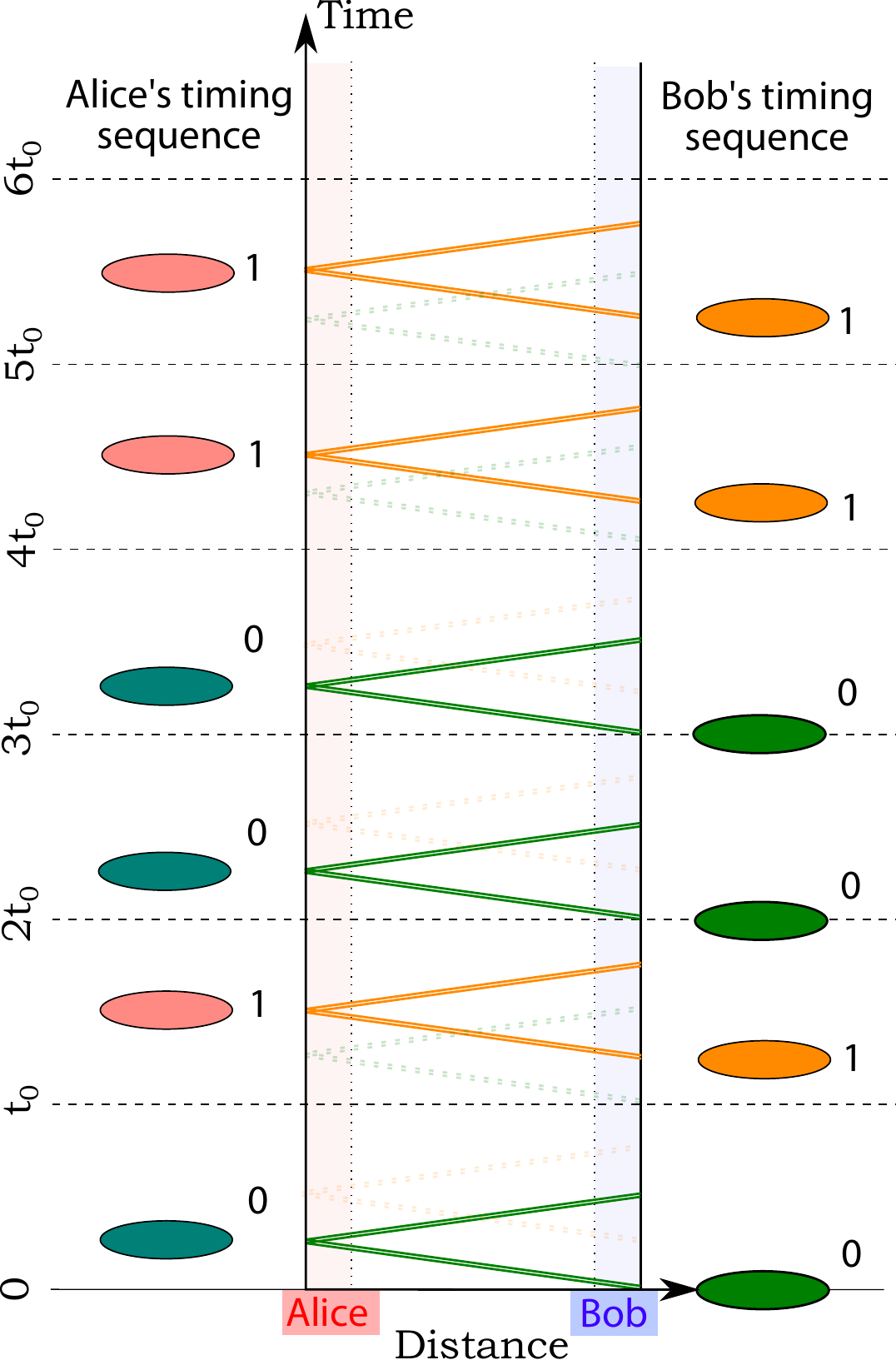}
\caption{{\bf 
Space-time diagram of a series of transmissions, realizing clock synchronization between Alice and Bob.} Initiating each
transmission, Bob randomly chooses whether to transmit at the beginning of a time slot of size $t_0$ or to transmit
delayed. This creates a unique Bob's timing sequence. Alice compares her observed timing sequence with that of Bob and
if they differ, Alice and Bob discard corresponding raw key sequence. As Bob's timing information reaches Alice with the
maximum possible speed --- the speed of light --- Eve's attempts to pre-poll Alice's random bit $b_A$ (see
figure~\ref{relativity}), keeping the
timing sequence the same, fail due to inability of faster than light delivery of timing sequence from Bob to Alice.
}\label{timing} 
\end{figure}

The second inherent part of the protocol guarantees clock synchronization between Alice and Bob. If it was missing, Eve
could use a fake pair of pulses to poll Alice's bit before getting a real signal from Bob. In this case Eve would always
know whether she was successful with the last measurement by the time she needs to make a decision about the first
state, which results in successful cracking of the protocol.  To avoid that Bob sends his pulses aperiodically: in each
data clock cycle he randomly chooses between the two time positions to send the pulse as shown in figure~\ref{timing}. This additional bit of
information cannot reach Alice earlier than it does under normal system operation. Any attempts of Eve to pre-poll
Alice's bits will result in a different time sequence observed by Alice, because to keep it, transmission of the timing
data from Bob to Alice would have been done superluminally.  According to the protocol, after series of transmissions
Alice and Bob compare their recorded timing data, taken by their unsynchronized, but precise in relative measurement
clocks, and if they observe errors in the timing sequence, they discard the whole series.

\begin{figure}
\centering
\includegraphics[width=\columnwidth]{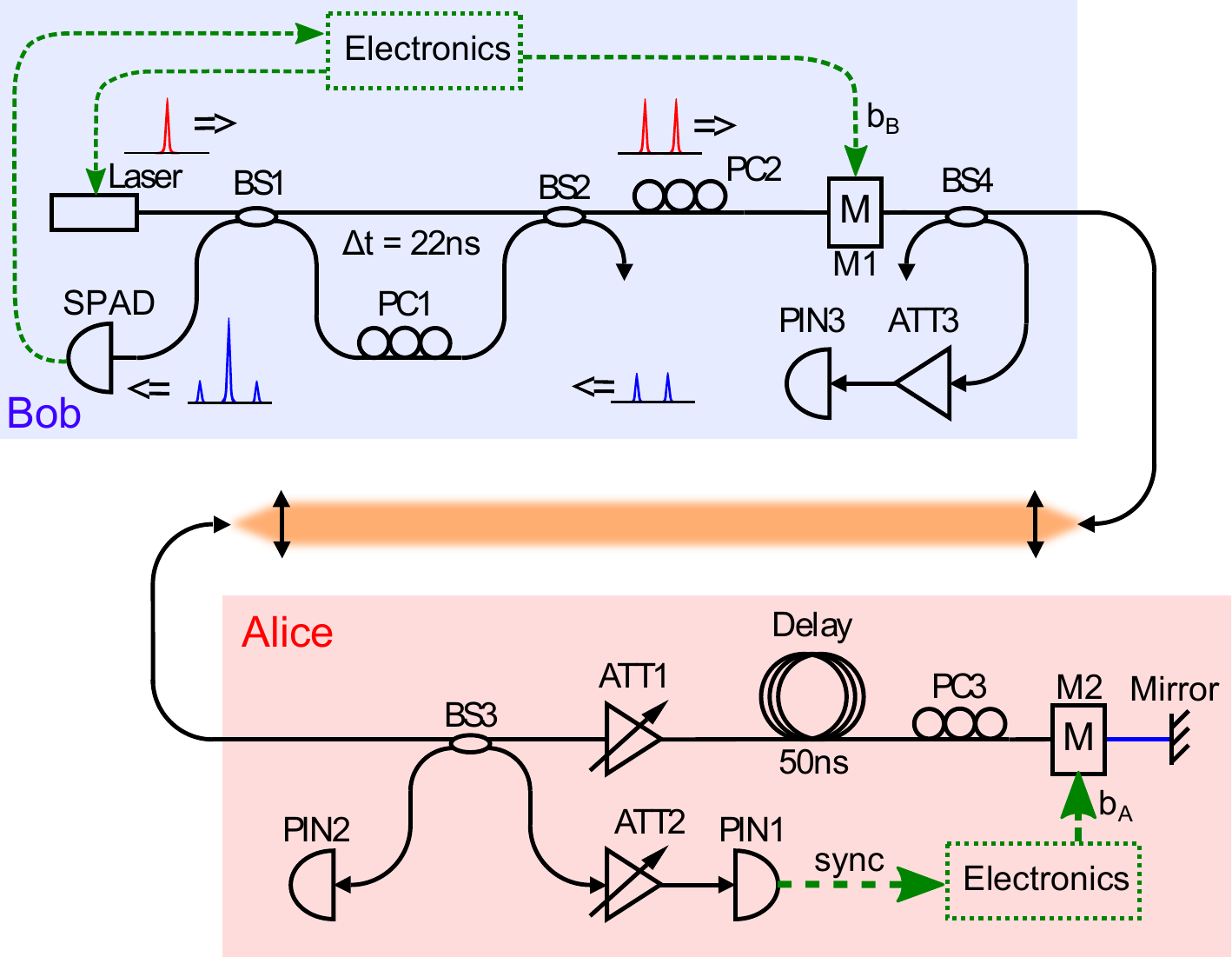}
{\caption{{\bf Experimental setup.} SPAD -- single photon avalanche detector, BS -- beam splitter, PC -- polarization controller,
M -- phase modulator, PIN -- PIN photodiode, ATT - attenuator. The setup consists of two fiber optic modules, Alice and Bob, and a
free-space optical channel between them. Phase-time encoding and decoding is performed in the same fiber delay interferometer located at
the Bob's side. Detection of photons by the SPAD
is performed in the middle time slot, which corresponds to interference between the two halves of the transmitted pulse.}\label{Exp_scheme}}
\end{figure}

Having realized both clock synchronization and key distribution in a single setup, we arrive at a relativistic protocol, where unlike
the conventional QKD, optical loss and errors caused by the eavesdropper are completely decoupled. Thus, even at an
arbitrarily large channel loss and unlimited Eve's resources, any extra information obtained by Eve results in
an increase of detection errors observed by Bob and detection of the intrusion.

\begin{figure}
\centering
\includegraphics[width=\columnwidth]{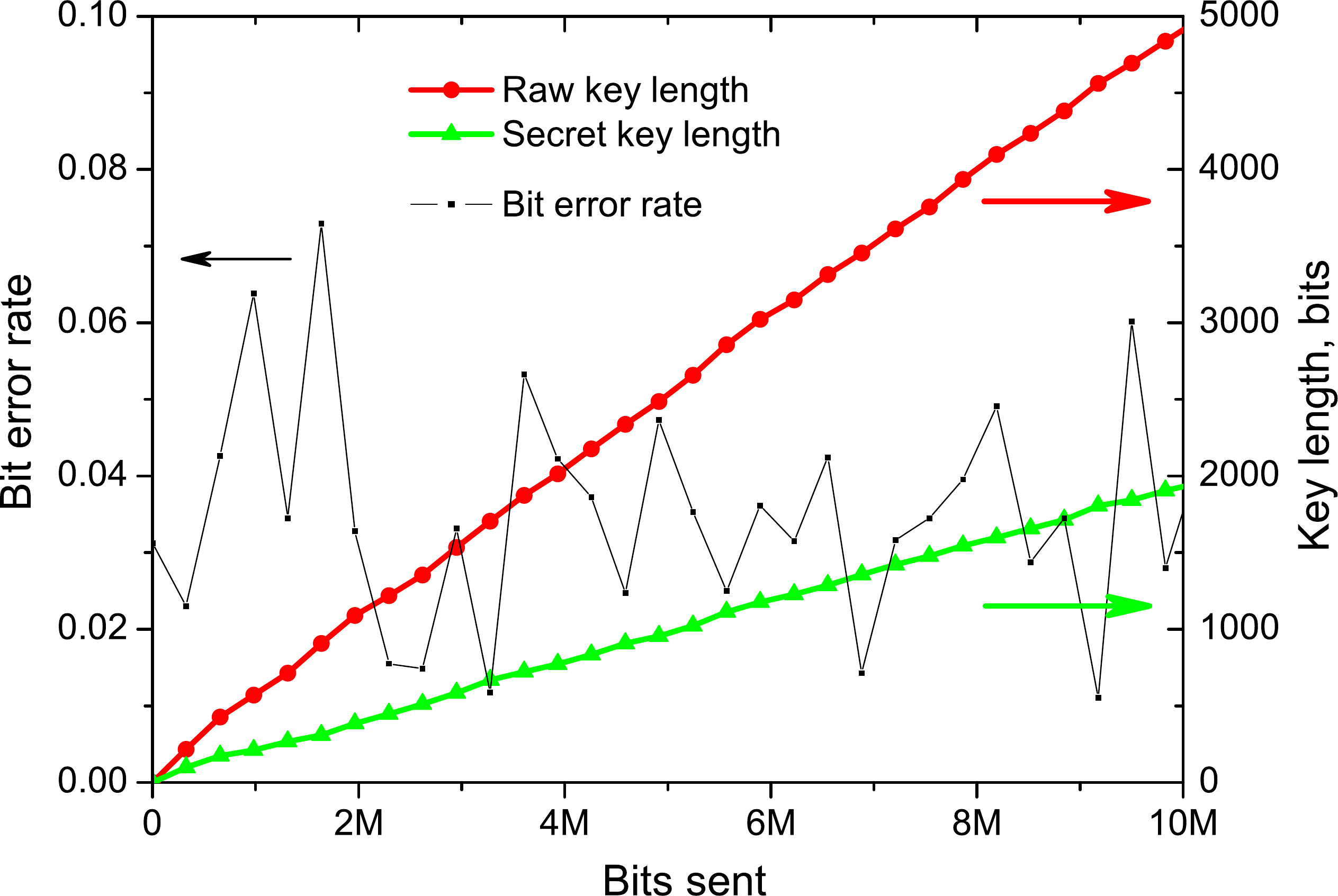}
\caption{{\bf Experimentally measured bit error rate and the obtained key lengths.} 
During the run over 55~m long free space channel Alice was checking her observed timing
sequence and compared it with that used by Bob. No timing errors were observed.
Average number of photons per
modulated pulse was kept at $\mu = 0.1$ and a depth of phase modulation was equal $130^\circ.$ Detection of arriving photons
was performed by Bob in a 4-ns time window, which is 5.5 times less than $\Delta t = 22$~ns, satisfying the
requirements of the relativistic protocol.
}\label{results} 
\end{figure}

The realized experimental system (figure~\ref{Exp_scheme}) is an optical fiber-based setup working at the wavelength of 850 nm with a 55~m long free-space channel
between Alice and Bob (see Appendix B: Experimental setup.) In this demonstration with the clock
rate of 250~kHz and $\mu = 0.1$~photons per pulse we obtained the average of 16.1~raw bits per series of 32768 pulses
with a 3.5\%~quantum bit error ratio as shown in figure~\ref{results}. This corresponds to a raw key generation rate of 123~bits/s and the
secret key rate of 47~bits/s.

To conclude, we have proposed and experimentally demonstrated a novel type of QKD protocols based on 
the principle of relativistic causality. Security of such a scheme with a conventional faint laser photon source does
not depend on a loss in the quantum channel, so it can be readily used for unconditionally secure satellite-based QKD networking,
with the only practical reach limit in the presence of dark photodetector counts.

\begin{acknowledgments}
This work was supported in part by the Russian Ministry of Education
and Science (state contracts no. 11.519.11.4009, 14.132.21.1400, 16.740.11.0662) and by RFBR, grant
no. 12-02-31792.
\end{acknowledgments}

\appendix
\section{Secret key rate.}
Here we present a brief asymptotic analysis
of the protocol performance, while a comprehensive study including finite sequences can be found in~\cite{M12}.
The secret key rate is bounded by
$$
R = \lim_{n\rightarrow\infty}
\frac{l_{secr}}{n} 
\le
(1-\eta)
(1-C(\varphi))-\eta-h(p_e)
,
$$
where $\eta$ is the fraction of errors in the received timing sequence (assumed to be zero throughout the paper),
$h(x)$ --- binary entropy function, $p_e$ --- bit error probability, and $C(\varphi)$ --- Holevo bound~\cite{H98} on classical throughput of the quantum channel
with states $|\alpha\rangle$ and $|e^{i\varphi}\alpha\rangle$;
$C(\varphi) = h\left(\frac{1-\varepsilon}{2}\right),$ where $\varepsilon = |\langle\alpha|e^{i\varphi}\alpha\rangle| = \exp\left(-2\mu\sin^2(\varphi/2)\right)$.

Typically in the experiments (when there is no active eavesdropping) there are no errors in the received timing sequences,
therefore we have chosen a simple strategy of discarding all pakets with timing errors. In this case
$\eta=0$ and the above expression becomes $R\le1-C(\varphi)-h(p_e)$, which has a simple intuitive interpretation:
if we assume that all classical information that could be derived from the sent quantum sequence is known to Eve, from each raw
bit we should subtract Eve's information $C(\varphi)$ and the entropy associated with bit errors in the raw sequence $h(p_e)$.

One can also notice that there is an ambiguity of choosing $\mu$ and $\varphi$ for a particular value of $C(\varphi)$.
However, from a practical point of view it is convenient to keep $\varphi$ close to $\pi$ to minimize the effect of experimental
errors.
In the particular experimental realization $C(\varphi) = 0.387$~bits, and the secret key rate depends on the observed $p_e$.
Importantly, if the rate $R$ becomes zero or even negative, as in the case of a large $\mu$ and a non-zero
$p_e$, there is no private shared information between Alice and Bob, i.e. all the raw key bits obtained have to be discarded.

\section{Experimental setup.} To generate optical pulses we use a directly modulated Fabri-Perot laser diode (QPhotonics
QFLD-850-75S), which emits 4~ns long pulses at the wavelength of 850~nm. The delay interferometer is fusion spliced from
a single-mode fiber HP780 and comprises a polarization controller (General Photonics PCD-M02-4X-NC-4) in one of the
arms. We use lithium-niobate phase modulators (Photline NIR-MPX800-LN-05) and mechanical variable optical attenuators (OZ
Optics DD-100-11-850-5/125-S-50) for further light processing. The transmission line is constructed from a pair of
free-space couplers (Micro Laser Systems, Inc. FC20-NIR-T) with an output aperture of 23~mm placed on tripods; the channel
loss is estimated as 3~dB. The setup on the other side of the free-space channel (Alice) is made of the same components
and works in a slave mode. A signal from the PIN1 photodetector activates the setup, which selectively modulates the
second optical pulse after its reflection from the mirror (OZ Optics FORF-11P-850-5/125-P). All random number generators in the
whole system (two of them control corresponding phase modulators and one creates a timing sequence) are emulated by a pair of
linear feedback shift register-based pseudo random generators with a sequence length of $2^{20}-1$ working synchronously on the both sides of
the channel. Thus, this proof-of-principle setup does not require a classical communication channel.
Single-photon detection is performed with a thermoelectrically cooled actively quenched avalanche photodiode (Excelitas C30902SH),
having a 30\% detection efficiency.

\section{Single-mode free-space channel requirements.}
An ideal single-mode free-space channel is described by a paraxial wave equation with a solution in the form
of a Gaussian beam. Beam diffraction limits the maximal length of the line, which becomes dependent on the
lens diameter. In a general symmetric configuration the channel length
$$L=\frac{2\pi w^2}\lambda \frac {w_0}w\sqrt{1-\left(\frac {w_0}w\right)^2},$$
where $w_0$ is the beam waist, and $w$ -- radius of the beam at the lens. The length is maximized at
$w_0/w = 1/\sqrt{2},$ as shown in Fig.~\ref{supp1}{\bf a.} In this configuration
$$L = \frac{\pi w^2}{\lambda} = 2\frac{\pi w_0^2}{\lambda}= 2z_R,$$
where $z_R$ is the Rayleigh length.

In our current implementation with 23~mm diameter free-space couplers, the radius $w$ equals 5.8~mm, resulting in
the working transmission distance up to $L \approx 125$~m. 
The demonstrated distance of 55~m, thus, should be considered
as a proof-of-principle demonstration limited by the length of the hallway, rather than the actual range of the system.
Larger, kilometer range, distances may be covered with larger free-space couplers
or telescopes. As optical loss in such free-space lines strongly depends on atmospheric turbulence
and fluctuates with time, the key generation rate may vary drastically, but without compromising system security due to
the unique properties of the {\it relativistic} quantum key distribution.

\section{Hardware implementation.}
\begin{figure}
\centering
\includegraphics[width=0.8\columnwidth]{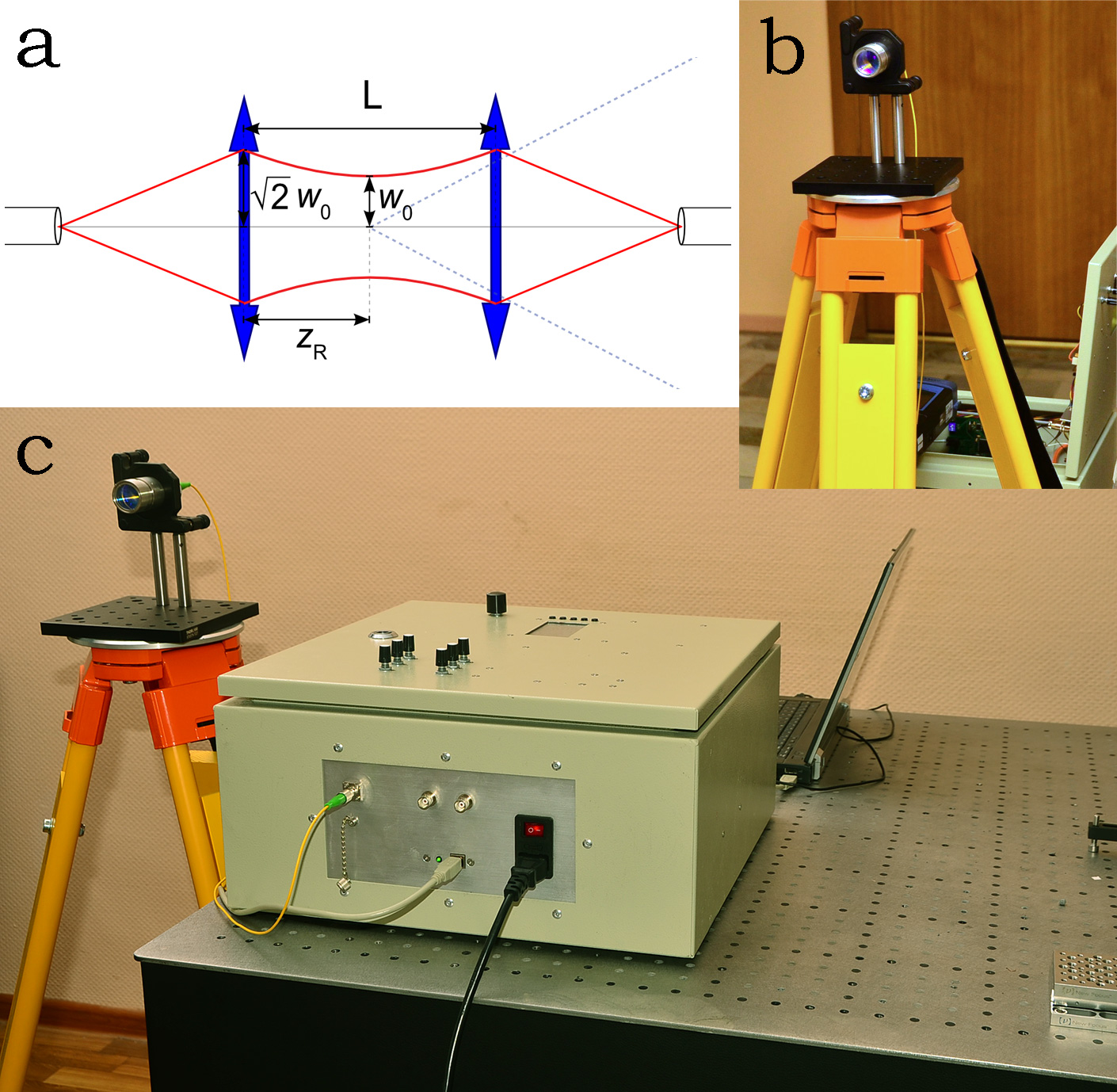}
\caption{{\bf Experimental realization.} {\bf a,} Optimal configuration of the free-space channel, achieving the
longest transmission distance $L$; $w_0$ -- beam waist, $z_R$ -- Rayleigh length. The figure shows two single-mode
fiber tips and two lenses forming a Gaussian beam. Horizontal dimensions are not to scale. {\bf b,} 'Alice' station with a
free-space coupler on a tripod. {\bf c,} 'Bob' station comprising a free-space coupler,
the station itself, and a laptop computer for data collection and analysis.
}\label{supp1} 
\end{figure}

Overall, the system consists of two boxes housing Alice and Bob, and a pair of tripods with the free-space couplers mounted on them, see Fig.~\ref{supp1}{\bf b,c.}
The boxes contain no free-space optical components so all light processing is realized in single-mode fiber-optic elements.
Each box is connected to a free-space coupler via a single-mode fiber and may be connected to a computer with a USB interface.

\begin{figure}
\centering
\includegraphics[width=\columnwidth]{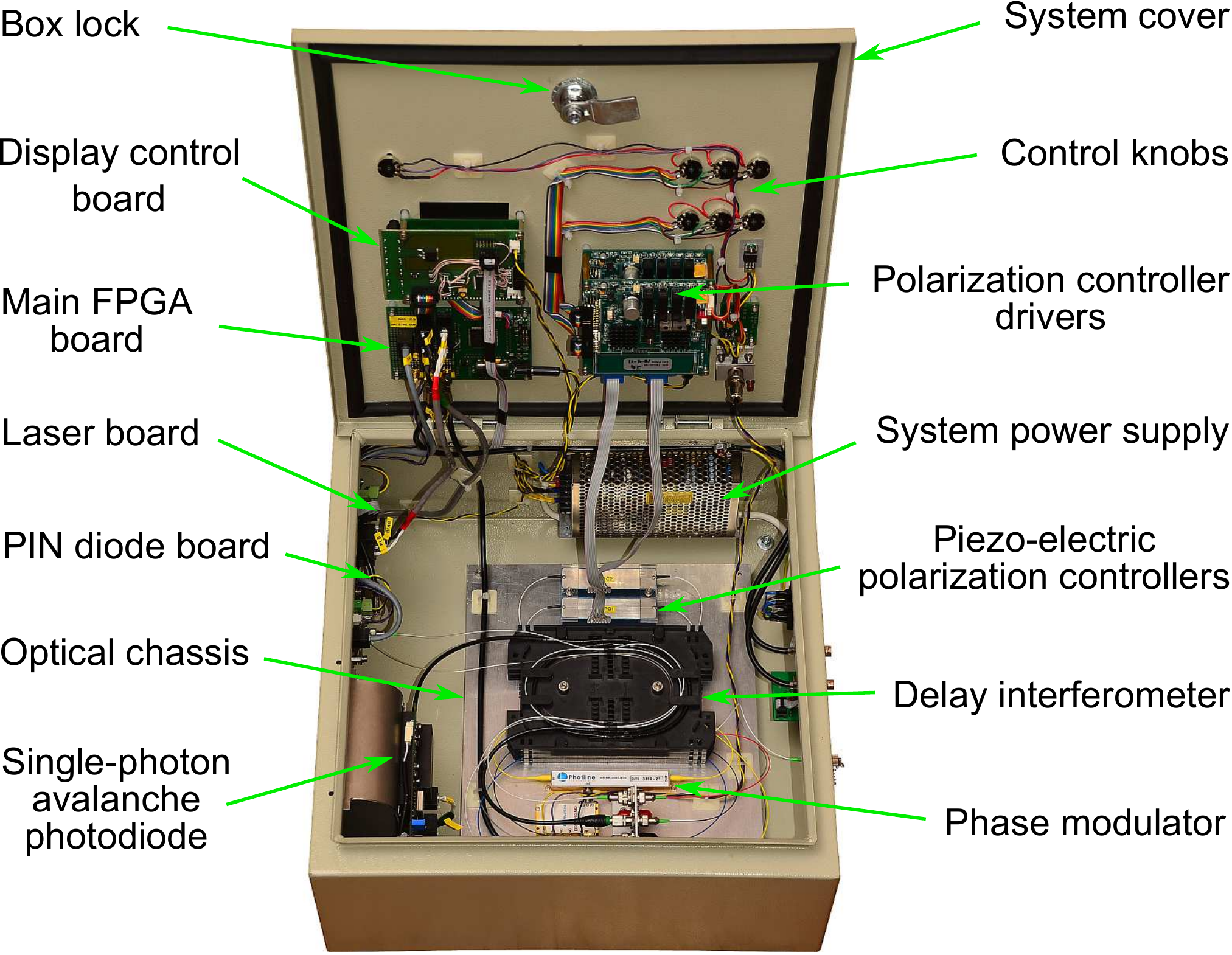}
\caption{{\bf Hardware implementation of the Bob's station.} The station is packaged in a metal box with a cover, which
has control knobs, buttons, and a small LCD display for visualization of the main operation parameters. It
connects to a computer via a USB cable for transmission of the obtained raw keys as well as for exchange of control information.
}\label{supp2} 
\end{figure}

All electronics controlling generation, processing and detection of optical signals is packaged in the same boxes side by side
with optical components as shown in Fig.~\ref{supp2}. High-speed electronic functions are directly performed by a
field-programmable gate array (FPGA), which forms the core of each box. Auxiliary functions such as computer connectivity, display control,
etc. are performed in a microcontroller.

\section{Comparison with previous relativistic approaches.}
The first {\it relativistic} QKD protocol was proposed in 1995~\cite{GV95}, where due to relativistic causality it was
possible to use orthogonal quantum states. As discussed in~\cite{P96,GV96}, the actual states in the channel are not
orthogonal because each state as a whole is never present in the channel. 
This approach can be classified as a whole class of relativistic protocols
with quantum states spread for more than the channel
length. Interestingly, the same idea of only partial quantum state presence in
the channel, later found another application in counterfactual QKD~\cite{N09,RWW11}.

Following the first pioneer work, in~\cite{KI97} it was shown that transmission at unknown random moments of time is not required if
globally non-orthogonal quantum states are used instead. 
Even with this modification, the protocol remained quite impractical as it requires on-site  phase-stable delays equal to
the time-of-flight between Alice and Bob and also precise clock synchronization between the parties.
Nevertheless, it has been recently realized as a table-top experiment~\cite{ABD10}
and even as a 1 km long fiber realization~\cite{XTW12}. Further theoretical developments~\cite{CS14} allowed to use
shorter time delays and to obtain more than one secret key bit per photon transmitted, however, the requirement of a true
single photon source and a lossless channel remained.

The next major breakthrough in relativistic quantum cryptography happened with the invention of relativistic bit
commitment protocols~\cite{K99,MN00}. Without relativity principles, as has been shown earlier by Mayers, Lo, and Chau,
quantum bit commitment is impossible~\cite{M96,M97,LC97}. This is probably the first example of substantially new
quantum cryptography protocol enabled by adding the theory of relativity. Later a few other relativistic bit commitment protocols have
been proposed~\cite{H11,K12} and even demonstrated experimentally~\cite{LKB13}.

The present paper shows a particular realization of a fundamentally new relativistic QKD approach, which uses
relativistic principles to strengthen conventional QKD protocols. In result, the new relativistic protocols demonstrate
unconditional security of generated keys, regardless of the quantum channel loss and with classical photon sources.
It has to be mentioned that the proposed principle can be used as a general framework for future relativistic QKD
protocols, whose application range is much superior than for conventional, non-relativistic QKD.


\begin{thebibliography}{10}
\providecommand{\url}[1]{#1}
\csname url@samestyle\endcsname
\providecommand{\newblock}{\relax}
\providecommand{\bibinfo}[2]{#2}
\providecommand{\BIBentrySTDinterwordspacing}{\spaceskip=0pt\relax}
\providecommand{\BIBentryALTinterwordstretchfactor}{4}
\providecommand{\BIBentryALTinterwordspacing}{\spaceskip=\fontdimen2\font plus
\BIBentryALTinterwordstretchfactor\fontdimen3\font minus
  \fontdimen4\font\relax}
\providecommand{\BIBforeignlanguage}[2]{{%
\expandafter\ifx\csname l@#1\endcsname\relax
\typeout{** WARNING: IEEEtran.bst: No hyphenation pattern has been}%
\typeout{** loaded for the language `#1'. Using the pattern for}%
\typeout{** the default language instead.}%
\else
\language=\csname l@#1\endcsname
\fi
#2}}
\providecommand{\BIBdecl}{\relax}
\BIBdecl

\bibitem{BB84}
C.~H. Bennett and G.~Brassard, ``Quantum cryptography: Public key distribution
  and coin tossing,'' in \emph{Proceedings of the IEEE International Conference
  on Computers, Systems, and Signal Processing}, Bangalore, 1984, pp. 175--179.

\bibitem{E91}
A.~K. Ekert, ``Quantum cryptography based on {B}ell's theorem,'' \emph{Phys.
  Rev. Lett.}, vol.~67, no.~6, pp. 661--663, 1991.

\bibitem{GRT02}
N.~Gisin, G.~Ribordy, W.~Tittel, and H.~Zbinden, ``Quantum cryptography,''
  \emph{Rev. Mod. Phys.}, vol.~74, no.~1, pp. 145--195, 2002.

\bibitem{SBC09}
V.~Scarani, H.~Bechmann-Pasquinucci, N.~J. Cerf, M.~Du\v{s}ek,
  N.~L{\"u}tkenhaus, and M.~Peev, ``The security of practical quantum key
  distribution,'' \emph{Rev. Mod. Phys.}, vol.~81, no.~3, pp. 1301--1350, 2009.

\bibitem{HH11}
R.~Hughes and J.~Nordholt, ``Refining quantum cryptography,'' \emph{Science},
  vol. 333, no. 6049, pp. 1584--1586, 2011.

\bibitem{LR13}
P.~K. Lam and T.~C. Ralph, ``Quantum cryptography: Continuous improvement,''
  \emph{Nature Photon.}, vol.~7, no.~5, pp. 350--352, 2013.

\bibitem{LM00}
B.~Lounis and W.~E. Moerner, ``Single photons on demand from a single molecule
  at room temperature,'' \emph{Nature}, vol. 407, no. 6803, pp. 491--493, 2000.

\bibitem{B00}
S.~Benjamin, ``Quantum cryptography - single photons 'on demand',''
  \emph{Science}, vol. 290, no. 5500, p. 2273, 2000.

\bibitem{D82}
D.~Dieks, ``Communication by {EPR} devices,'' \emph{Phys. Lett. A}, vol.~92,
  no.~6, pp. 271--272, 1982.

\bibitem{WZ82}
W.~K. Wootters and W.~H. Zurek, ``A single quantum cannot be cloned,''
  \emph{Nature}, vol. 299, no. 5886, pp. 802--803, 1982.

\bibitem{GV95}
L.~Goldenberg and L.~Vaidman, ``Quantum cryptography based on orthogonal
  states,'' \emph{Phys. Rev. Lett.}, vol.~75, no.~7, pp. 1239--1243, 1995.

\bibitem{P96}
A.~Peres, ``Quantum cryptography with orthogonal states?'' \emph{Phys. Rev.
  Lett.}, vol.~77, no.~15, p. 3264, 1996.

\bibitem{GV96}
L.~Goldenberg and L.~Vaidman, ``Goldenberg and vaidman reply,'' \emph{Phys.
  Rev. Lett.}, vol.~77, no.~15, p. 3265, 1996.

\bibitem{KI97}
M.~Koashi and N.~Imoto, ``Quantum cryptography based on split transmission of
  one-bit information in two steps,'' \emph{Phys. Rev. Lett.}, vol.~79, no.~12,
  pp. 2383--2386, 1997.

\bibitem{ABD10}
A.~Avella, G.~Brida, I.~P. Degiovanni, M.~Genovese, M.~Gramegna, and P.~Traina,
  ``Experimental quantum cryptography scheme based on orthogonal states,''
  \emph{Phys. Rev. A}, vol.~82, p. 062309, 2010.

\bibitem{XTW12}
G.~B. Xavier, G.~P. Temporao, and J.~P. von~der Weid, ``Employing long
  fibre-optical {M}ach-{Z}ehnder interferometers for quantum cryptography with
  orthogonal states,'' \emph{Electron. Lett}, vol.~48, no.~13, pp. 775--777,
  2012.

\bibitem{JAK06}
E.~Jeffrey, J.~Altepeter, and P.~G. Kwiat, ``Relativistic quantum
  cryptography,'' in \emph{OSA Frontiers in Optics}, Rochester, New York, Oct.
  2006.

\bibitem{K99}
A.~Kent, ``Unconditionally secure bit commitment,'' \emph{Phys. Rev. Lett.},
  vol.~83, no.~7, pp. 1447--1450, 1999.

\bibitem{MN00}
S.~N. Molotkov and S.~S. Nazin, ``Relativistic quantum bit commitment in
  real-time,'' \emph{JETP}, vol.~90, no.~4, pp. 714--723, 2000.

\bibitem{H11}
G.~P. He, ``Quantum key distribution based on orthogonal states allows secure
  quantum bit commitment,'' \emph{J. Phys. A: Math. Theor.}, vol.~44, p.
  445305, 2011.

\bibitem{K12}
A.~Kent, ``Unconditionally secure bit commitment by transmitting measurement
  outcomes,'' \emph{Phys. Rev. Lett.}, vol. 109, p. 130501, 2012.

\bibitem{LKB13}
T.~Lunghi, J.~Kaniewski, F.~Bussi\'{e}res, R.~Houlmann, M.~Tomamichel, A.~Kent,
  N.~Gisin, S.~Wehner, and H.~Zbinden, ``Experimental bit commitment based on
  quantum communication and special relativity,'' \emph{Phys. Rev. Lett.}, vol.
  111, no.~18, p. 180504, 2013.

\bibitem{M96}
D.~Mayers, ``The trouble with quantum bit commitment,'' 1996,
  arXiv:quant-ph/9603015.

\bibitem{M97}
------, ``Unconditionally secure quantum bit commitment is impossible,''
  \emph{Phys. Rev. Lett.}, vol.~78, no.~17, p. 3414, 1997.

\bibitem{LC97}
H.~K. Lo and H.~F. Chau, ``Is quantum bit commitment really possible?''
  \emph{Phys. Rev. Lett.}, vol.~78, no.~17, p. 3410, 1997.

\bibitem{MN01}
S.~N. Molotkov and S.~S. Nazin, ``The role of causality in ensuring the
  ultimate security of relativistic quantum cryptography,'' \emph{JETP Lett.},
  vol.~73, no.~12, pp. 682--687, 2001.

\bibitem{M11}
S.~N. Molotkov, ``Relativistic quantum cryptography for open space without
  clock synchronization on the receiver and transmitter sides,'' \emph{JETP
  Lett.}, vol.~94, no.~6, pp. 469--476, 2011.

\bibitem{M11_VAR}
------, ``Relativistic quantum cryptography,'' \emph{JETP}, vol. 112, no.~3,
  pp. 370--379, 2011.

\bibitem{M12}
------, ``On the resistance of relativistic quantum cryptography in open space
  at finite resources,'' \emph{JETP Lett.}, vol.~96, no.~5, pp. 342--348, 2012.

\bibitem{B92}
C.~H. Bennett, ``Quantum cryptography using any two nonorthogonal states,''
  \emph{Phys. Rev. Lett.}, vol.~68, no.~21, pp. 3121--3124, 1992.

\bibitem{H98}
A.~S. Holevo, ``Quantum coding theorems,'' \emph{Russian Math. Surveys},
  vol.~53, no.~6, pp. 1295--1331, 1998.

\bibitem{N09}
T.-G. Noh, ``Counterfactual quantum cryptography,'' \emph{Phys. Rev. Lett.},
  vol. 103, no.~23, p. 230501, 2009.

\bibitem{RWW11}
M.~Ren, G.~Wu, E.~Wu, and H.~Zeng, ``Experimental demonstration of
  counterfactual quantum key distribution,'' \emph{Laser Physics}, vol.~21,
  no.~4, pp. 755--760, 2011.

\bibitem{CS14}
J.~S. Cotler and P.~W. Shor, ``A new relativistic orthogonal states quantum key
  distribution protocol,'' 2014, arXiv:quant-ph/1401.5493.

\end{thebibliography}
\end{document}